\documentclass[twocolumn,prX,showpacs]{revtex4}

\usepackage{amssymb,amsmath}
\usepackage{openbib}
\usepackage{epsfig}
\usepackage{amsthm}
\usepackage{epsfig}
\usepackage{subeqnarray}
\usepackage{graphicx}
\usepackage{floatrow}

\usepackage{notoccite} 
\usepackage{hyperref}  

\input{Qcircuit}

\newcommand{\eq}[1]{{eq.(\ref{#1})}}
\renewcommand{\theequation}{{\arabic{section}.\arabic{equation}}}  
\DeclareMathOperator{\lcm}{lcm}
\newfloatcommand{capbtabbox}{table}[][] 
\DeclareMathOperator{\Tr}{Tr}

\begin{document}

\title{Simplified Factoring Algorithms for Validating Small-Scale Quantum Information Processing Technologies}

\author{Omar Gamel}\email{ogamel@physics.utoronto.ca}\author{Daniel F. V. James} 
\address{Department of Physics, University of Toronto, 60 St. George St., Toronto, Ontario, Canada M5S 1A7.}

\date{November 14, 2013}

\begin{abstract}
Tomography has reached its practical limits in characterization of new quantum devices, and there is a need for a new means of characterizing and validating new technological advances in this field. We propose a different verification scheme based on compiled versions of Shor's factoring algorithm that may be extended to large circuits in the future. The general version Shor's algorithm has been experimentally elusive due to bottlenecks associated with the modular exponentiation operation. Experiments to date have only been able to execute compiled versions of the latter operation. We provide some new compiled circuits for experimentalists to use in the near future. We also demonstrate that an additional layer of compilation can be added using classical operations, that will reduce the number of qubits and gates needed in a given compiled circuit.
\end{abstract}


\maketitle

\renewcommand{\theequation}{\arabic{equation}}

\section{Introduction}

Theory remains very much ahead of experiment in quantum computing. While complicated quantum algorithms are developed and studied, experimental capabilities are not advanced enough to implement them. As experimental techniques progress on the road to a fully functional quantum computer, there is a great need for small scale algorithms and circuit verification techniques to verify the functionality of new technology. 

Traditionally, tomography has been used to characterize and validate quantum computing technology, and test the functionality of new devices \cite{munrojames, robin, jamesKaz}. However, despite advances in compressed sensing and other enhancements, full quantum state tomography beyond more than one or two dozen qubits appears to be intractable due to the amount of measurements required (which grows exponentially with the number of qubits) \cite{flammia}. Therefore, it is of interest to create new methods of verifying that new quantum devices actually execute their desired function.

To this end, it is natural to use well known algorithms, such as Shor's factoring algorithm \cite{shor}, for verification and characterization of new technology. However, the complete version of Shor's algorithm uses an order finding subroutine that involves a modular exponentiation operation, which is the bottleneck of the algorithm, needing the most quantum gates to implement and the most time to execute. Creating circuits for a general modular exponentiation operation is not possible with currently realizable technology for any nontrivial parameters. 

There has been much theoretical work on construction of quantum circuits, with efficient techniques developed to simplify circuit construction \cite{beckman,vedral,vanmeter, saeedi, saeedi2}. Among other uses, these techniques have been employed to construct ``compiled circuits" for the order-finding subroutine in Shor's algorithm and associated modular exponentiation operation. Compilation in this sense uses known information about the solution to simplify the circuit from its complicated general form to a more manageable size. Additionally, experimental demonstrations of compiled circuits have been carried out using various technologies \cite{lanyon, lu, martin, obrien, vandersypen, xu, lucero}. Experimental implementation of a non-compiled version of Shor's algorithm may indeed soon be forthcoming \cite{blatt2}. 

In this paper we will use compiled versions of the modular exponentiation operation for the purposes of validation. We demonstrate some techniques to construct the simple compiled circuits, including a ``classical layer" of compilation that reduces the number of qubits needed. It remains to be seen if one can generalize these compilation and validation techniques beyond a dozen or so qubits, to the regime where tomography has significant limitations.

In section \ref{sec2}, we provide a brief overview of Shor's algorithm and the steps it involves, distilling it to the modular exponentiation operation. Section \ref{sec3} demonstrates the process of constructing the compiled modular exponentiation circuits for some semiprimes, using circuit synthesis methods similar to those in ref. \cite{gamelPeriodic}. In section \ref{sec4} we briefly discuss the relationship between allowable periods in the modular exponentiation subroutine for each number to be factored. Section \ref{sec5} analyzes a specific example of compiled modular exponentiation and order-finding, taking into account the effects of noise and entanglement in the compiled circuits. 

\section{Shor's Algorithm}\label{sec2}

\subsection{Classical Probabilistic Steps}

Shor's factoring algorithm is held as one of the most promising and useful applications of quantum computing. It allows one to factor large numbers in polynomial time, undermining the most common cryptographic scheme in use today, RSA cryptography \cite{rsa}. The algorithm is based on an order-finding subroutine, which in turn makes heavy use of the modular exponentiation operation \cite{nielsenchuang, kaye}. The modular exponentiation is the bottleneck of the algorithm, needing the resources in terms of quantum gates and computation time. 
  
The purpose of Shor's algorithm is to factorize a number $N$. It is assumed $N$ is odd, and that it is not a power of a prime number. The case $N=p^k$ for some prime $p$ and integer $k$ can be checked by calculating $\sqrt[k]{N}$ (the $k^{th}$ root of $N$) for all integer $k \le \log_3{N}$, which is an efficient procedure \cite{nielsenchuang, kaye}. Since $N$  is not the power of a prime, it can be written as the product of two coprime integers. 

Shor's algorithm non-deterministically finds a nontrivial factor of $N$ in the following manner. First, a number $a < N$ is chosen arbitrarily. The greatest common divisor of $a$ and $N$, denoted $\gcd(a, N)$, is found using the Euclidean algorithm \cite{euclid}. If $\gcd(a, N) \ne 1$, then $\gcd(a, N)$ is a nontrivial factor of N, and no further work is needed. In case $a$ and $N$ are coprime, i.e. $\gcd(a, N) = 1$, the algorithm uses the order finding subroutine explained in the next subsection to find the order of $a$ modulo $N$. Defining the function $f_{a,N}(x)$ as
\begin{equation}
 f_{a,N}(x) \equiv a^x \mod (N),
\label{faNdef}
\end{equation}
the order of $a$ modulo $N$ is defined as the smallest positive integer $x$ such that  $f_{a,N}(x) = 1$. We denote the order $r$. The order $r$ is also the period of the function $f_{a,N}(x)$, that is $f_{a,N}(x)=f_{a,N}(x+r)$. For Shor's algorithm to succeed, either $r$ has to be even, or $a$ has to be a square of an integer, so that either way $a^{r/2}$ is an integer.

If $r$ is found to be odd, and $a$ is not a square, then the algorithm trial has failed and a new random number $a$ above must be attempted. If $r$ is even, we compute $a^{r/2} \mod (N)$. If  $a^{r/2} \equiv -1 \mod (N)$, once again the algorithm has failed and a new random number $a$ must be tried. We note that $a^{r/2} \ne 1 \mod (N)$, because if $a^{r/2}$ was congruent to $1$ modulo $N$ then $r$ would not be the smallest power that satisfies the order condition. 


Supposing the algorithm succeeded, we have that $a^{r/2} \ne -1 \mod (N)$ and $a^{r/2} \ne 1 \mod (N)$, yet $a^{r} = 1 \mod (N)$. These statements imply that $N \nmid (a^{r/2} + 1)$ and $N \nmid (a^{r/2} - 1)$, yet $N | (a^{r} - 1) =  (a^{r/2} + 1)(a^{r/2} - 1) $. We have used standard number theory notation, where $b|c$ means $b$ is a factor of $c$, and $b\nmid c$ means $b$ is not a factor of $c$. 

Since $N$ divides the product of $(a^{r/2} + 1)$ and $(a^{r/2} - 1)$, but not each of them alone, then the nontrivial factors of $N$ must be split between the two numbers. To find the factors, we simply use the Euclidean algorithm to calculate $\gcd(a^{r/2} \pm 1, N)$. With this, Shor's algorithm concludes.

Note that the processes involved in Shor's algorithm can all be executed efficiently (in polynomial time) on a classical computer, with the exception of order finding. It is a bottleneck for classical computing, since it requires a large number of evaluations of the underlying function; quantum parallelism removes this problem and fewer evaluations are needed, leading to an efficient computation. 

However, despite this, the order finding subroutine comes with its own limitations. Order finding itself is broken down to two steps; modular exponentiation and a quantum Fourier transform. The modular exponentiation is what requires the most resources. The quantum Fourier transform requires few resources, particularly when implemented semi-classically \cite{griffithsniu}. So Shor's algorithm while efficient, is still very difficult to apply practically given today's limitations in quantum technology. To see this, we discuss the order finding subroutine separately below.

\subsection{Quantum Order Finding Subroutine}
To find the order of the function $f_{a,N}(x) = a^x \mod (N)$ defined in \eq{faNdef}, Shor's algorithm makes heavy use of the modular exponentiation operation given by
\begin{equation}
U(a):\ket{x}\ket{0} \rightarrow \ket{x}\ket{a^x mod (N)}.
\label{modexp}
\end{equation}
Clearly the structure of the circuit that implements this operation will depend on the random number $a$ chosen. The quantum state is initialized at
\begin{equation}
\frac{1}{\sqrt{M}}\sum_{x=0}^M \ket{x}\ket{0},
\end{equation}
where $M=2^m$ is the smallest power of $2$ that satisfies $M \ge N^2$. We then apply the modular exponentiation operation in \eq{modexp} to our initial state, yielding
\begin{align}
\frac{1}{\sqrt{M}} U(a) \sum_{x=0}^M \ket{x}\ket{0} &= \frac{1}{\sqrt{M}} \sum_{x=0}^M \ket{x}\ket{a^x(mod N)}.
\end{align}
Applying the inverse quantum Fourier transform will yield a quantum state which when measured, will with a high probability lead us to the order $r$ of the function $f_{a,N}(x) = a^x \mod (N)$. 

Since the modular exponentiation operation $U(a)$ depends on the arbitrary number $a$, the general circuit implementing it must have $a$ as a variable input, making it very complicated. In other words, it is very difficult to create a single circuit that will execute the modular exponentiation in \eq{modexp} for arbitrary values of $x$ and $a$. For this reason, the modular exponentiation is by far the most resource intensive part of Shor's algorithm, and is currently the biggest practical obstacle to implementing the algorithm for useful values of $N$ \cite{saeedi2}. Though some experimental groups may soon realize an uncompiled Shor's algorithm \cite{blatt2}, they will still be limited to small numbers for the foreseeable future. 

To handle this difficulty while still implementing most aspects of Shor's algorithm, researchers have resorted to `compiled' versions of the algorithm. Practically, this means that knowing $N$ in advance, a specific value of the number $a$ is chosen that is known to yield a successful result within the algorithm. Then the circuit $U(a)$ for the modular exponentiation for the specific value of $a$ is constructed and used for the compiled algorithm. This will be much simpler than the generic algorithm, since we no longer need to construct a modular exponentiation circuit that works for any value of $a$, but for a single value.

Of course, this roundabout process does not really factor the number $N$, in fact it assumes previous knowledge of the solution. 
This has led some authors to suggest that ``to even call such a procedure compilation is an abuse of language" \cite{jsmolin}, since compilers cannot know the solution of the problem beforehand. 

Nonetheless, the so called compilation process is of interest as a proof of concept of aspects of the algorithm, and is often the best that can be done with current technology. It is of interest to explore the process of compiling circuits to create simple examples that provide milestones for experimentalists as technology advances.

\section{Compiled Circuits}\label{sec3}

\subsection{Implementations of Compiled Shor's Algorithm }
Much theoretical work has been completed on compiled circuits for the order-finding subroutine in Shor's algorithm. Beckman et. al. published one of the first theoretical works on the topic \cite{beckman}, discussing a possible implementation through ion traps. The focus of that work was minimizing the number of memory qubits and operations needed by using known properties about the number $N$ to be factored. In particular, $N=15$ was used as an example, and general methods developed for any $N$. 

Similar work in developing theoretical techniqes to compile the modular exponentiation subroutine was carried out by Vedral \cite{vedral}, Van Meter and Itoh \cite{vanmeter}, and most recently by Markov and Saeedi \cite{saeedi, saeedi2}. 

Additionally, experimental demonstrations of compiled circuits through photonic qubits \cite{lanyon, lu, obrien} as well those with recycled photonic qubits \cite{martin}, nuclear magnetic resonance \cite{vandersypen, xu}, and superconducting qubits \cite{lucero} have been carried out. 

As discussed above, compilation in the sense used in this field does not really simplify the problem at hand. Its main purpose is to give  milestones that closely resemble useful tasks to nascent experimental devices. This will help sharpen experimental techniques on the way to a full implementation of the algorithm. Moreover, Shor's algorithm involves several steps, compilation oversimplifies only one of them, leaving the other steps intact. Therefore, implementing a compiled version will directly help us improve the technique involved in these steps.

\subsection{The Compilation Process}


As explained above, compilation hitherto used consists of choosing a specific number $a$ that is known to factor a given $N$ via Shor's algorithm, and creating the circuit for the function $U(a)$ defined in \eq{modexp}. We will demonstrate compilation examples for the semiprimes $N=15, 21,$ and $33$. These are the smallest numbers that are the product of two distinct odd primes. 

The modular exponentiation function is periodic, and injective within a single period. Therefore, we will use methods of circuit synthesis similar to those used in construction of simple periodic functions which have the same properties, discussed in ref. \cite{gamelPeriodic}. In the process, we show we can use a simple classical function to reduce the number of qubits and gates needed. One may call this a `layer of classical compilation'.

The circuits in this paper are constructed by inspection, and trial and error, making use of some general patterns and principles. Each circuit can be seen as two processes. The first is copying a linear combination of the input qubits to each output qubit via CNOT gates. This process must be used to create linearly independent combinations of the input bits in the output bits, which serve as the canvas on which the second process will operate. The linear independence ensures the condition of injectivity within a single period is satisfied. 

The second process is the application of cascades of Toffoli gates to modify the results of the first process by flipping some entries in the truth table. The first Toffoli gate of the cascade uses two suitable control bits. The Toffoli gate at each subsequent level of the cascade uses as one of its control bits the target bit of the previous Toffoli gate. It is this Toffoli cascade process where most of the creativity lies, since it is what actually creates the desired periodicity. The first process alone cannot create odd periodicity. Note that the two processes may be made to occur in tandem. Also note that all output qubits $y_i$ are initialized to the state $\ket{0}$.

If one thinks of the truth tables below, the first Toffoli gate applied modifies a number of entries equal to a quarter of the total length of the column (i.e. $2^{n-2}$). Each consecutive Toffoli gate in the cascade flips half the number of entries of its target qubit as in the previous level of the cascade. For example, if $n=3$ (as is the case for $S_5$ and $S_7$), then the first Toffoli gate will flip $2^{3-2} = 2$ entries, the second gate in the cascade will flip 1 entry. If $n=4$, then the first Toffoli gate will flip $2^{4-2} = 4$ entries, the second gate in the cascade will flip 2 entries, and the third gate will flip 1 entry. For $n$ input qubits, a Toffoli cascade will have $n-1$ gates, with the final gate flipping only 1 entry in the truth table.

Where possible, the output of a Toffoli cascade is copied over to other qubits via CNOT gates, to avoid need for an identical cascade. This copying may take place in the middle of, or at the end of the cascade. Some Toffoli gates may not be part of a cascade at all. 

We use standard notation for CNOT and Toffoli gates, with a black circle indicating the control qubit, and a large circle with a plus sign inside indicates the target qubit. The small white circles we see in some circuits (such as fig. \ref{circuitcompiledhalff4_21}) are inverted control qubits, in the sense that the target bit is modified if the inverted control bit has value $0$, and is left unchanged if it has value $1$.

The underlined entries in the truth tables denote the entries flipped by the action of a Toffoli gate, whether directly or indirectly (where the result of the Toffoli is copied by a CNOT to another bit). Each underline is an entry flip, so an even number of underlines leaves the entry unchanged.

\subsubsection{N=15}
We begin by considering the case $N=15$. As in ref. \cite{lanyon}, we consider two subcases, $a=2$ resulting in $r=4$, and $a=4$ with $r=2$. Starting with the case $a=2$, we have table \ref{tablef2_15} showing the truth table for $y=f_{2,15}(x)$, for input $x$ between $0$ and $3$. We use standard binary notation, and the input $x$ is represented by 2 qubits $x_2$, and $x_1$ and the output $y$ is represented by 4 qubits $y_4, y_3, y_2, $ and $y_1$. For example, if $x=2$, then $x_2 = 1$, and $x_1 = 0$. 

\begin{table}[H] \centering
\begin{tabular}{l l | r r r r }  
$x_2$ & $x_1$ & $y_4$ & $y_3$ & $y_2$ & $y_1$ \\ 
  \hline \hline
   0 & 0 & 0 & 0 & 0 & 1 \\
   0 & 1 & 0 & 0 & 1 & 0 \\
   1 & 0 & 0 & 1 & 0 & 0 \\
   1 & 1 & $\underline{1}$ & $\underline{0}$& $\underline{0}$ & $\underline{0}$\\
\end{tabular}
\caption{\label{tablef2_15}The binary truth table for $y=f_{2,15}(x)$.}
\end{table}

The function $y=f_{2,15}(x)$ can be implemented using the circuit in fig. \ref{circuitf2_15} below. The underlined entries in the truth table above are the ones that were modified by the action of a Toffoli gate, whether directly or indirectly via copying through a CNOT gate. We also reuse the definition of $N_T$ as the number of Toffoli gates in a circuit and $N_{CN}$ as the number of CNOT gates.
\begin{figure}[H] \centering
\
\Qcircuit @C=1em @R=.7em { 
\lstick{x_2} 	& \ctrlt{1} 	& \ctrl{5} 	& \qw	 & \qw 	& \qw 	& \qw 	& \qw 	& \ctrl{3}	& \qw	& \rstick{x_2}  \\
\lstick{x_1} 	& \ctrlt{1} 	& \qw 	& \ctrl{4} 	& \qw 	& \qw 	& \qw 	& \ctrl{3}	& \qw 	& \qw	& \rstick{x_1}  \\
\lstick{\ket{0}}  	& \targc	& \qwc	& \qwc	& \ctrlc{3}	& \ctrlc{2}	& \ctrlc{1}	& \qwc	& \qwc	& \qwc	& \rstick{y_4}  \\
\lstick{\ket{0}}  	& \qwc	& \qwc	& \qwc	& \qwc	& \qwc	& \targc	& \qwc	& \targc	& \qwc	& \rstick{y_3}  \\
\lstick{\ket{0}}  	& \qwc	& \qwc	& \qwc	& \qwc	& \targc	& \qwc	& \targc	& \qwc	& \qwc	& \rstick{y_2}  \\
\lstick{\ket{0}}  	& \targc	& \targc	& \targc	& \targc	& \qwc	& \qwc	& \qwc	& \qwc	& \qwc	& \rstick{y_1}  \\
}
\caption{\label{circuitf2_15}The circuit for $y=f_{2,15}(x)$. Gate count: $N_T=1$ and $N_{CN}$ = 7. This is slightly cheaper in terms of gates than the analogous arrangement in ref. \cite{lanyon}, which used 2 controlled-swap gates (roughly as hard as a Toffoli) and 2 CNOT gates.}
\end{figure}
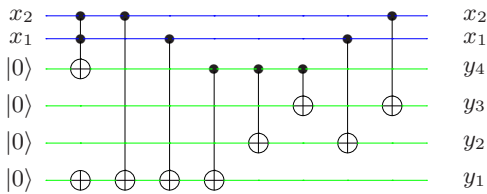

The authors in ref. \cite{lanyon} define a fully compiled function, which we denote $\tilde{f}_{2,15}(x) \equiv \log_2(f_{2,15}(x))$. For $x=0$ to $3$, we have $\tilde{f}_{2,15}(x)=x$. The circuit for this function is much simpler than the one above, and is given in fig. \ref{circuitcompiledf2_15} below. Although this circuit is a gross oversimplification of modular exponentiation, it is still in some sense ``compiled".
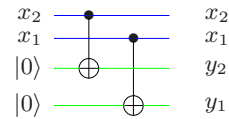
\begin{figure}[H] \centering
\ 
\Qcircuit @C=1em @R=.7em { 
\lstick{x_2} 	& \ctrl{2} 	& \qw 	& \qw	& \rstick{x_2}  \\
\lstick{x_1} 	& \qw 	& \ctrl{2} 	& \qw 	& \rstick{x_1} \\
\lstick{\ket{0}} 	& \targc 	& \qwc 	& \qwc	&  \rstick{y_2}  \\
\lstick{\ket{0}} 	& \qwc 	& \targc 	& \qwc	&  \rstick{y_1}  
}
\caption{\label{circuitcompiledf2_15}The fully compiled circuit for $y=\tilde{f}_{2,15}(x)$. Gate count: $N_T=0$, $N_{CN}$= 2. }
\end{figure}

The use of the $\log_2$ function can be seen as a layer of ``classical compilation". We exploit this idea later in the paper to generalize it to other sorts of functions.

Moving to the case $a=4$, the truth table for $y=f_{4,15}(x)$, for input $x$ equal to $0$ or $1$ is in table \ref{tablef4_15}.  

\begin{table}[H] \centering
\begin{tabular}{l | r r r }  
$x_1$ & $y_3$ & $y_2$ & $y_1$ \\ 
  \hline \hline
   0 & 0 & 0 & 1  \\
   1 & 1 & 0 & 0 
\end{tabular}
\caption{\label{tablef4_15}The binary truth table for $y=f_{4,15}(x)$.}
\end{table}

The circuit for $y=f_{4,15}(x)$ is given in fig. \ref{circuitf4_15} below.

\begin{figure}[H] \centering
\ 
\Qcircuit @C=1em @R=.7em { 
\lstick{x_1} 		& \ctrl{1} 	& \ctrl{3}	& \qw		& \rstick{x_1}  \\
\lstick{\ket{0}}	& \targc	& \qwc 	& \qwc	& \rstick{y_3} \\
\lstick{\ket{0}} 	& \qwc 	& \qwc 	& \qwc	&  \rstick{y_2}  \\
\lstick{\ket{0}} 	& \targc 	& \targc 	& \qwc	&  \rstick{y_1}  
}
\caption{\label{circuitf4_15}The circuit for $y=f_{4,15}(x)$. Gate count: $N_T$=0, $N_{CN}$ = 2.}
\end{figure}
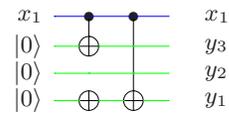

\begin{figure}[H] \centering
\ 
\Qcircuit @C=1em @R=.7em { 
\lstick{x_1} 		& \ctrl{1} 	& \qw		& \rstick{x_1}  \\
\lstick{\ket{0}} 	& \targc 	& \qwc	&  \rstick{y_1}  
}
\caption{\label{circuitcompiledf4_15}The fully compiled circuit for $y=\tilde{f}_{4,15}(x)$. Gate count: $N_T$=0, $N_{CN}$ = 1.}
\end{figure}
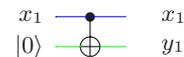

Single qubit gates such as the NOT gate are easy to implement, and we do not factor them into the gate count. Although this circuit is simple as it is, one can fully compile it further \cite{lanyon} by defining $\tilde{f}_{4,15}(x) \equiv \log_4(f_{4,15}(x))$. This function simply maps $0$ and $1$ to themselves, and can be implemented through a single CNOT gate as in fig. \ref{circuitcompiledf4_15}.

The above compilation for $N=15$ has been implemented in ref. \cite{lanyon}. We generalize these ideas and drive them further for larger $N$. We will construct the general uncompiled circuits, and then compile them through different intermediate functions to reach simpler circuits.

\subsubsection{N=21}
Suppose we wish to compile the factorization of $N=21$, and that we have chosen $a=4$ for this purpose. To get a better intuitive understanding of the modular exponentiation process, we compute the values of $f_{4,21}(x) = 4^x \mod(21)$. The results are in table \ref{tabledecimalf4_21} below.

\begin{table}[H] \centering
\begin{tabular}{c | c c c c c c c c c c c }  
   $x$ 			& 0 & 1 &  2  & 3   & 4   & 5   & 6 & 7 \\ \hline
   $4^x \mod(21)$ 	& 1 & 4 & 16 & 1   & 4   & 16 & 1 & 4 \\ 
\end{tabular}
\caption{\label{tabledecimalf4_21}The decimal value table for $f_{4,21}(x)$.}
\end{table}
From the values in the table, we note that $f_{4,21}(x)$ is has period $3$. Therefore, modulo $21$, the order of $4$ is $3$.
Similarly, for any $a$ coprime to $21$ one can find the corresponding order $r$ modulo $21$. We do so in the table \ref{ordersmod21}. 

\begin{table}[H] \centering
\begin{tabular}{c | c c c c c c c c c c c }  
   $a$ & 2 & 4 & 5 & 8 & 10 & 11 & 13 & 16 & 17 & 19 & 20 \\ \hline
   $r$ & 6 & 3 & 6 & 2 & 6 & 6 & 2 & 3 & 6 & 6 & 2 \\
\end{tabular}
\caption{\label{ordersmod21}The period $r$ of $f_{a,21}(x)$ for all $a$ coprime to 21.}
\end{table}

Note that all the periods are factors of $6$, which is to be expected for $N=21$ as will be explained in section \ref{sec4}. For now, we simply note that the largest period is $6$, meaning that in general, to implement $U(a)$ defined in \eq{modexp}, one needs at most 3 qubits in the input register. Returning to the function $y=f_{4,21}(x)$, we include its truth table for input $x$ between $0$ and $7$ in table \ref{tablef4_21}.

\begin{table}[H] \centering
\begin{tabular}{l l l | r r r r r }  
$x_3$ & $x_2$ & $x_1$ & $y_5$& $y_4$ & $y_3$ & $y_2$ & $y_1$ \\ 
  \hline \hline
   0 & 0 & 0 & 0 & 0 & 0 & 0 & 1\\
   0 & 0 & 1 & 0 & 0 & 1 & 0 & 0 \\
   0 & 1 & 0 & $\underline{1}$ & 0 &  $\underline{0}$ & 0 & 0 \\
   0 & 1 & 1 & 0 & 0 & 0 & 0 & 1 \\
   1 & 0 & 0 & 0 & 0 & 1 & 0 & 0 \\
   1 & 0 & 1 & $\underline{1}$ & 0 &  $\underline{\underline{0}}$ & 0 &  $\underline{0}$ \\
   1 & 1 & 0 & 0 & 0 & 0 & 0 & 1 \\
   1 & 1 & 1 & 0 & 0 & 1 & 0 & 0 \\
\end{tabular}
\caption{\label{tablef4_21}The binary truth table for $y=f_{4,21}(x)$.}
\end{table}


We make some observations about table \ref{tablef4_21}, first that $y_4$ and $y_2$ are always zero, and need not have any gates. We observe that $y_3$ can \emph{almost} be written as $y_3 = x_3 \oplus x_2 \oplus x_1$, with the exception being the underlined $0$ in the third row, which would have been $1$ if the formula was exact. 

Similarly $y_1$ can \emph{almost} be written as $y_1 = x_3 \oplus x_2 \oplus x_1 \oplus 1$, with the exception being the underlined $0$ in the sixth row, which would have been a $1$ if the formula was exact. 

We also note that $y_5$ has only two nonzero entries, which can potentially be constructed from a single Toffoli gate. Observe that $y_5$ takes the value $1$ when $x_3=x_1\ne x_2$, or equivalently when $x_3 \oplus x_2 = x_2 \oplus x_1 = 1$. Therefore, we can construct $y_5$ as the output of a Toffoli gate with the two control bit values $x_3 \oplus x_2$ and $x_2 \oplus x_1$. 

Then we can use another Toffoli gate (forming a Toffoli cascade) that has $y_5$ as one control bit and $x_3$ as the other control bit to modify the entry in the fifth row of $y_1$, the target bit. To flip the entry in the third row for $y_3$, we simply add the result of the two previous Toffoli gates, that is, we add $y_5$ and the modifier entry of $y_1$, which will result in a single modifier entry in the third row of $y_3$. 

All of the above is shown in the circuit constructed in fig. \ref{circuitf4_21} below. 

\begin{figure}[H] \centering
\ 
\Qcircuit @C=.5em @R=.7em { 
\lstick{x_3} & \qw 	& \qw  	& \qw		& \ctrl{7}	& \qw 		& \ctrl{7}	& \qw 		& \ctrl{3}	& \qw		& \qw		& \ctrl{5}	& \qw		& \qw		& \ctrl{7}	& \qw		& \rstick{x_3}    \\ 
\lstick{x_2} & \ctrl{4} 	& \qw  	& \ctrl{6} 	& \qw  	& \qw 		& \qw 		& \ctrl{6} 	& \qw	 	& \qw		& \qw		& \qw		& \qw		& \ctrl{6}	& \qw		& \qw		& \rstick{x_2}  \\
\lstick{x_1} & \qw	& \ctrl{3}	& \qw 		& \qw  	& \qw 		& \qw 		& \qw 		& \qw		& \qw 		& \qw		& \qw		& \ctrl{5}	& \qw		& \qw		& \qw		& \rstick{x_1} \\
\lstick{\ket{0}} & \qwc 	& \qwc 	& \qwc 	& \qwc 	& \targc	& \qwc	& \qwc 	& \ctrlc{4} 	& \ctrlc{2}	& \qwc	& \qwc	& \qwc	& \qwc	& \qwc	& \qwc	& \rstick{y_5}  \\
\lstick{\ket{0}} & \qwc 	& \qwc 	& \qwc 	& \qwc 	& \qwc 	& \qwc 	& \qwc 	& \qwc 	& \qwc	& \qwc	& \qwc	& \qwc	& \qwc	& \qwc	& \qwc	& \rstick{y_4} \\
\lstick{\ket{0}} & \targc  	& \targc	& \qwc 	& \qwc 	& \ctrlc{-2}	& \qwc	& \qwc 	& \qwc 	& \targc	& \targc	& \targc	& \qwc	& \qwc	& \qwc	& \qwc	& \rstick{y_3}  \\
\lstick{\ket{0}} & \qwc 	& \qwc 	& \qwc 	& \qwc 	& \qwc	& \qwc 	& \qwc 	& \qwc 	& \qwc	& \qwc	& \qwc	& \qwc	& \qwc	& \qwc	& \qwc	& \rstick{y_2}  \\
\lstick{\ket{0}} & \qwc  	& \qwc 	& \targc 	& \targc 	& \ctrlc{-2}	& \targc 	& \targc 	& \targc 	& \qwc	& \ctrlc{-2}	& \targc	& \targc	& \targc	& \targc	& \qwc	& \rstick{y_1}    
}
\caption{\label{circuitf4_21}The circuit for $y=f_{4,21}(x)$ with three input qubits, and five output qubits. Gate count: $N_T$=2, $N_{CN}$ = 12. }
\end{figure}
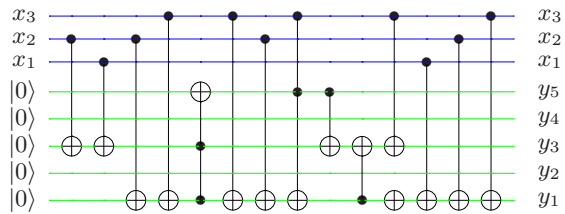

We have again underlined all entries in the truth table (table \ref{tablef4_21}) that are flipped by a Toffoli gate, whether directly or indirectly (i.e. the result of a Toffoli being copied by a CNOT). An entry that is flipped twice (i.e. unchanged) is doubly underlined.

This circuit synthesis procedure is ad-hoc, and based on taking advantage of some patterns in the function truth table. It is much more efficient than some alternatives ones. It uses only 2 Toffoli gates and 12 CNOT gates (with a quantum cost of 24), whereas for example, the circuit for the identical function in ref. \cite{saeedi} uses 8 Toffoli gates and 5 CNOT gates (quantum cost of 53).

%

Even so, we note that the circuit above has 5 output qubits to represent the three output values $1, 4,$ and $16$. To partially compile the circuit, we map these three outcomes to the integers $0,1,$ and $2$ instead, via the operation of a logarithm base $4$. 

That is, we define a function $\hat{f}_{4,21}(x) \equiv \log_4(f_{4,21}(x))$, which needs only two out put qubits. Its truth table is shown in table \ref{tablecompiledhalff4_21} below. Using our same circuit synthesis procedure, the circuit for $\hat{f}_{4,21}(x)$ is shown in fig. \ref{circuitcompiledhalff4_21}.
%
%
%
\begin{figure}[H] \centering
\ 
\Qcircuit @C=1em @R=.7em { 
\lstick{x_3} & \targ 	& \qw 		& \ctrl{2}	& \targ	& \ctrlo{3}	& \qw		& \ctrl{4}	& \qw		& \qw		& \rstick{x_3}  \\
\lstick{x_2} & \ctrl{-1} 	& \ctrl{1} 	& \qw		& \ctrl{-1} 	& \qw 		& \qw 		& \qw		& \ctrl{1} 	& \qw		&  \rstick{x_2} \\
\lstick{x_1} & \qw 	& \targ	& \ctrl{1} 	& \qw 		& \qw 		& \ctrl{2}	& \qw		& \targ	& \qw		&  \rstick{x_1}  \\
\lstick{\ket{0}} & \qwc 	& \qwc	& \targc	& \qwc	& \ctrlc{1}	& \qwc	& \qwc	& \qwc	& \qwc	&  \rstick{y_2}  \\
\lstick{\ket{0}} & \qwc 	& \qwc 	& \qwc 	& \qwc 	& \targc	& \targc 	& \targc	& \qwc	& \qwc	&  \rstick{y_1}  
}
\caption{\label{circuitcompiledhalff4_21} The circuit for the partially compiled $\hat{f}_{4,21}(x)$, with three input qubits and two output qubits. Gate count: $N_T$=2, $N_{CN}$ = 6.}
\end{figure}
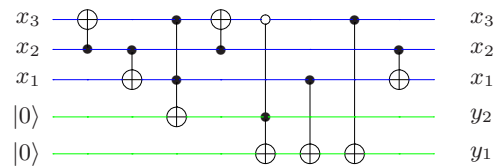
\begin{table}[H] \centering
\begin{tabular}{l l l | r r }  
$x_3$ & $x_2$ & $x_1$ & $y_2$& $y_1$ \\ 
  \hline \hline
   0 & 0 & 0 & 0 & 0 \\
   0 & 0 & 1 & 0 & 1 \\
   0 & 1 & 0 &  $\underline{1}$ &  $\underline{0}$ \\
   0 & 1 & 1 & 0 & 0 \\
   1 & 0 & 0 & 0 & 1 \\
   1 & 0 & 1 &  $\underline{1}$ & 0 \\
   1 & 1 & 0 & 0 & 0 \\
   1 & 1 & 1 & 0 & 1 \\
\end{tabular}
\caption{\label{tablecompiledhalff4_21}The truth table for the partially compiled $\tilde{f}_{4,21}(x)$, with three input qubits and two output qubits.}
\end{table}%
%
%
%
The function $\log_4$ can be used through an intermediate classical operation, and simplifies the circuit to 2 Toffoli gates and 6 CNOT gates (quantum cost of 18), and uses fewer qubits. Again, the logarithm can be considered a classical layer of compilation that reduces the number of qubits needed to more manageable levels.

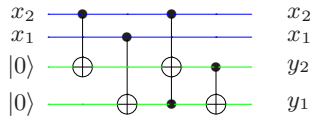
\begin{figure}[H] \centering
\ 
\Qcircuit @C=1em @R=.7em { 
\lstick{x_2} & \ctrl{2} 	& \qw 		& \ctrlt{2} 	& \qw 		& \qw		& \rstick{x_2}  \\
\lstick{x_1} & \qw 	& \ctrl{2} 	& \qw		& \qw 		& \qw 		& \rstick{x_1} \\
\lstick{\ket{0}} & \targc 	& \qwc 	& \targc 	& \ctrlc{1} 	& \qwc	&  \rstick{y_2}  \\
\lstick{\ket{0}} & \qwc 	& \targc 	& \ctrlct{-1} 	& \targc 	& \qwc	&  \rstick{y_1}  
}
\caption{\label{circuitcompiledf4_21} The fully compiled circuit for $y=\tilde{f}_{4,21}(x)$ with an additional layer of classical compilation. Gate count: $N_T$=1, $N_{CN}$ = 3. Identical to $S_3$ circuit in ref. \cite{gamelPeriodic}.}
\end{figure}

Since the circuit in fig. \ref{circuitcompiledhalff4_21} has period 3, it can be further compiled to the fully compiled function $\tilde{f}_{4,21}(x)$, whose truth table is in table \ref{tablecompiledf4_21} and circuit in fig. \ref{circuitcompiledf4_21}. This fully compiled circuit has only two input qubits. It can be thought of as the ``underlying circuit". Note that in a fully compiled circuit, the values of the function output generated by all the possible inputs create one full period of the function, but not two or more. This is similar to the idea of the simplest periodic function. In fact, the function $\tilde{f}_{4,21}(x)$ is identical to $S_3$, the simple monoperiodic function of period 3, introduced in ref. \cite{gamelPeriodic}. Many fully compiled circuits for higher $N$ will also fall into the simple periodic function category.
\begin{table}[H] \centering
\begin{tabular}{l l | r r }  
$x_2$ & $x_1$ & $y_2$& $y_1$ \\ 
  \hline \hline
   0 & 0 & 0 & 0 \\
   0 & 1 & 0 & 1 \\
   1 & 0 &  $\underline{1}$ &  $\underline{0}$ \\
   1 & 1 & 0 & 0 
\end{tabular}
\caption{\label{tablecompiledf4_21}The fully compiled truth table for $y=\tilde{f}_{4,21}(x)$ with two input and two output qubits.}
\end{table}%
%
%
\subsubsection{N=33}

Suppose $N=33$, the table \ref{ordermod33} shows all the possible $a$ coprime to $33$, and the order of each modulo $33$ (i.e. the period of $f_{a,33}(x)$). 

\begin{table}[H] \centering
\scalebox{0.95}{\begin{tabular}{c | c c c c c c c c c c c c c c c c c c c}  
$a$ & 2  & 4 & 5   & 7   & 8   & 10 & 13 & 14 & 16 & 17 & 19 & 20 & 23 & 25 & 26 & 28 & 29 & 31 & 32 \\ \hline
$r$ & 10 & 5 & 10 & 10 & 10 & 2   & 10 & 10 & 5  & 10  & 10 & 10 & 2   & 5   & 10 & 10 & 10 & 5  & 2
\end{tabular}}
\caption{\label{ordermod33}The period $r$ of $f_{a,33}(x)$ for all $a$ coprime to 33.}
\end{table} 

We choose $a=4$ once again, since it is easy to work with, and its odd period of $5$ is admissible in Shor's algorithm since $4$ is a square. We can construct the circuit for calculating $f_{4,33}(x) = 4^x \mod (33)$, which takes on values $1, 4, 16, 31, 25$ as $x$ varies from $0$ to $4$, and then repeats for $x\ge 5$. To represent the input values $0$ to $4$ we need three qubits in the input register, and to represent the output values up to $31$ we will need five qubits in the output register.

However, we choose instead to construct the compiled circuit, for $\tilde{f}_{4,33}(x) = g(f_{4,33}(x))$ where $g(y)=(y-1)/3$. This maps the output values to $0, 1, 5, 10, 8$. This map works well since $g(y)$ is a simple differentiable function that maps the output values of $f_{4,33}(x)$ to smaller integer values. Here, $g(y)$ performs the classical compilation step, and takes the place of the logarithms used in previous examples. Constructing the circuit for $g(f_{4,33}(x))$ we have fig. \ref{circuitf4_33} below.

\begin{figure}[H] \centering
\ 
\Qcircuit @C=1em @R=.7em { 
\lstick{x_3} & \qw 		& \targ	& \ctrl{1} 	& \targ 	& \qw 		& \qw 		& \ctrlo{2}	& \ctrl{3}	& \qw 		& \qw 		& \qw 		& \rstick{x_3}    \\ 
\lstick{x_2} & \targ 		& \qw 		& \ctrl{2}	& \qw 		& \targ 	& \ctrl{2}	& \qw 		& \qw 		& \ctrl{3}	& \ctrl{5}	& \qw 		& \rstick{x_2}    \\ 
\lstick{x_1} & \ctrl{-1}		& \ctrl{-2} 	& \qw 		& \ctrl{-2}	& \ctrl{-1} 	& \qw	 	& \ctrl{4}	& \qw 		& \qw 		& \qw 		& \qw 		& \rstick{x_1}    \\ 
\lstick{\ket{0}} & \qwc 		& \qwc 	& \targc 	& \qwc 	& \qwc 	& \ctrlc{1} 	& \qwc	& \targc 	& \qwc	& \qwc 	& \qwc	& \rstick{y_4}    \\ 
\lstick{\ket{0}} & \qwc 		& \qwc 	& \qwc 	& \qwc 	& \qwc 	& \targc 	& \qwc	& \qwc 	& \targc	& \qwc 	& \qwc	& \rstick{y_3}    \\ 
\lstick{\ket{0}} & \qwc		& \qwc 	& \qwc 	& \qwc 	& \qwc 	& \qwc 	& \qwc	& \qwc 	& \qwc	& \qwc 	& \qwc	& \rstick{y_2}    \\ 
\lstick{\ket{0}} & \qwc 		& \qwc 	& \qwc 	& \qwc 	& \qwc 	& \qwc 	& \targc	& \qwc 	& \qwc	& \targc 	& \qwc	& \rstick{y_1}    
}
\caption{\label{circuitf4_33}The circuit table for $y=\tilde{f}_{4,33}(x)$. Gate count: $N_T$=3, $N_{CN}$ = 7.}
\end{figure}
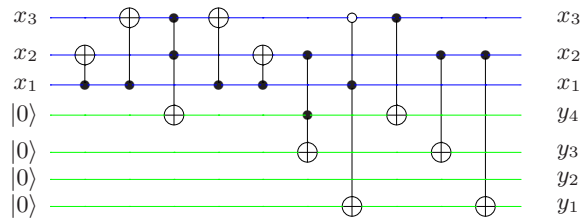

\begin{table}[H] \centering
\begin{tabular}{l l l | r r r r }  
$x_3$ & $x_2$ & $x_1$ & $y_4$ & $y_3$ & $y_2$ & $y_1$ \\ 
  \hline \hline
   0 & 0 & 0 & 0 & 0 & 0 & 0\\
   0 & 0 & 1 & $\underline{0}$ & 0 & 0 & $\underline{1}$ \\
   0 & 1 & 0 & 0 & 1 & 0 & 1 \\
   0 & 1 & 1 & 1 & 1 & 0 & $\underline{0}$ \\
   1 & 0 & 0 & 1 & 0 & 0 & 0 \\
   1 & 0 & 1 & 0 & 0 & 0 & 0 \\
   1 & 1 & 0 & $\underline{0}$ & $\underline{0}$ & 0 & 1 \\
   1 & 1 & 1 & 0 & 1 & 0 & 1 \\
\end{tabular}
\caption{\label{tablef4_33}The binary truth table for $y=\tilde{f}_{4,33}(x)$.}
\end{table}

\subsection{Classical Compilation}

%

What we have called the classical compilation step works by defining $\tilde{f}_{a,N}(x) = g(f_{a,N}(x))$ where $g(y)$ is some differentiable function that maps the output values of $f_{a,N}(x)$ (for all $x$ in the function's domain) to smaller integer values. The smaller output values after applying the function $g$ allow us to use fewer qubits to represent the output register, and thus saving us some resources. We have seen examples where $g(y)$ is a logarithm, or a linear function. In general, other simple functions can be used, though for some combinations of $a$ and $N$ this compilation is not possible except through very artificially defined complicated functions $g$. 

Circuit compilation is only interesting when it involves simple intermediate steps that offer a substantial simplification of a complex circuit. If the function we use to compile is itself overly complicated, then we have not gained much. This puts a limitation of the usefulness compilation process as a whole.

\section{General $N$ and Allowable Periods}\label{sec4} 
Given that the period is what determines the difficulty of factorization \cite{jsmolin}, and even the size of the compiled circuit, it is of interest to find which periods of $f_{a,N}(x) = a^x \mod(N)$ are allowed for a given $N$. That is, we wish to find the set of all orders of any $a$ modulo a given $N$.

Suppose $N=pq$ is fixed, where $p$ and $q$ are distinct primes. From elementary number theory we know there are $\varphi(N) = (p-1)(q-1)$ positive integers less than $N$ that are coprime to $N$ \cite{mermin}. The function $\varphi(N)$ is called Euler's totient function \cite{hardy}. The group formed by coprime integers modulo $N$ is of size $(p-1)(q-1)$, and so any subgroup must have a size, also known as the group's order, that divides $(p-1)(q-1)$. In other words, for any number $a$ chosen, the period of $f_{a,N}(x)$ must be a factor of $(p-1)(q-1)$. 

In fact, an even stronger condition has been established. The period $r$ of $f_{a,N}(x)$ can be shown to divide the Carmichael function of $N$ \cite{carmichael}, given by
\begin{equation}
\lambda(N) \equiv \lcm(p-1,q-1),
\label{carmichaelfunc}
\end{equation}
where $\lcm$ is the least common multiple.

For example, if $N=21$, then $p=3$ and $q=7$, therefore $\lambda(21) = \lcm(2,6) = 6.$ Table \ref{Ntable} gives pairs of distinct odd primes $p$, $q$, and the corresponding $N$, $\lambda(N)$, and allowed values of the period $r$ (which are all the factors of $\lambda(N)$) for $N < 90$.

\begin{table}[h] \centering
\begin{tabular}{c | c | c || c | l }  
$p$ & $q$ & $N$& $\lambda(N)$ & allowed $r$ \\ 
  \hline
   3 & 5 & 15 & 4 & 2, 4 \\
   3 & 7 & 21 & 6 & 2, 3, 6\\
   3 & 11 & 33 & 10 & 2, 5, 10 \\
   5 & 7 & 35 & 12 & 2, 3, 4, 6, 12 \\
   3 & 13 & 39 & 12 & 2, 3, 4, 6, 12 \\ 
   3 & 17 & 51 & 16 & 2, 4, 8, 16 \\
   5 & 11 & 55 & 20 & 2, 4, 5, 10, 20 \\
   3 & 19 & 57 & 18 & 2, 3, 6, 9, 18 \\
   5 & 13 & 65 & 12 & 2, 3, 4, 6, 12 \\
   3 & 23 & 69 & 22 & 2, 11, 22 \\
   7 & 11 & 77 & 30 & 2, 3, 5, 6, 10, 15, 30 \\
   5 & 17 & 85 & 16 & 2, 4, 8, 16 \\
   3 & 29 & 87 & 28 & 2, 4, 7, 14, 28 
\end{tabular}
\caption{\label{Ntable} For each semiprime $N=pq$ with $p$ and $q$ distinct odd primes, the table lists the value of the Carmichael function $\lambda(N) \equiv \lcm(p-1,q-1)$, and the allowable periods $r$, given by all the factors of $\lambda(N)$.}
\end{table}

Table \ref{Ntable} is useful for experimentalists who want to construct a compiled factoring circuit that has a given period/order. For example, if we wish to construct a compiled circuit for factoring some number $N$ where the period $r$ is $11$, then we know the smallest such $N$ is $69$.

\section{Illustrative Example of Compiled Period Finding}\label{sec5}
Suppose we are able to construct the compiled circuits described in the previous section using some quantum device. We can run the circuit with the initial state $\ket{+}$ in all input registers (generating an equal superposition), and $\ket{0}$ in all the output registers. After running the circuit we can measure the probability distribution of the output state and compare it to theoretical expectations to benchmark the device, and judge if it works correctly. In addition, by comparing the measured probability distribution to the theoretically expected one, we can roughly quantify the noise (assuming a Werner state model), and entanglement of the state. We demonstrate this by way of an illustrative example.

Suppose we have a ``toy" quantum circuit with three qubits in the input register and three in the output register. If we represent the binary numbers in each register as decimal numbers, the value in each register is an integer from $0$ to $7$. Suppose further that the circuit is periodic. That is, if we treat the circuit as a function with this input range, it has a period $p$, for some $p$ between 1 and 8 inclusive. 

Denoting the action of the circuit by the function $F_p$, and initializing each of the three qubits in the input register to the state $\ket{+}$, then the action of the circuit creates the output state $\ket{\psi_{p}}$ given by
\begin{align}
\ket{\psi_{p}} &\equiv F_p\Big[\frac{1}{\sqrt{8}}\sum_{j=0}^7 \ket{j}_i \ket{0}_o\Big] \nonumber\\
&= \frac{1}{\sqrt{8}}\sum_{j=0}^7 \ket{j}_i \ket{j \  \text{mod} p}_o,
\label{psip}
\end{align}
where $i$ subscript denotes the input register and $o$ subscript denotes the output registers.
For example, if $p=3$, we have
\begin{align}
\ket{\psi_{3}} = \frac{1}{\sqrt{8}} \Big[ &\big(\ket{0}+ \ket{3}+\ket{6} \big)_i\ket{0}_o + \big(\ket{1}+ \ket{4}+\ket{7} \big)_i\ket{1}_o \nonumber\\ 
&+ \big(\ket{2}+ \ket{5} \big)_i\ket{2}_o \Big].
\label{psi3}
\end{align}
As a side note, the state $\ket{\psi_{3}}$ is the one produced by the partially compiled circuit for $\hat{f}_{4,21}$ in fig. \ref{circuitcompiledhalff4_21}  and table \ref{tablecompiledhalff4_21}, with the subtle difference the here we have defined three qubits in the output register, and the $\hat{f}_{4,21}$ only has two.

In Shor's algorithm, the final step is applying the quantum Fourier transform (QFT) to the input register. We follow suit and do the same in our toy model. The QFT is defined as
\begin{equation}
\text{QFT}: \ket{j} \rightarrow \frac{1}{\sqrt{N}}\sum_{k=0}^{N-1} \omega_{N}^{kj} \ket{k},
\end{equation}
where $\omega_N$ is the $N$th root of unity, defined as $\omega_N \equiv e^{2\pi i/N}$. Continuing with our example, we have $N=2^3=8$. Dropping the subscript, we have $\omega\equiv \omega_8=e^{\pi i/4}=\frac{1}{\sqrt{2}}(1+i)$. 
%
%
Applying the the QFT to the input register in \eq{psip}, we have
\begin{equation}
\ket{\phi_p} \equiv \text{QFT}_i\ket{\psi_p} = \frac{1}{8}\sum_{j=0}^7\sum_{k=0}^{7} \omega^{kj} \ket{k}_i \ket{j \  \text{mod} p}_o. 
\label{phip}
\end{equation}
For example, if $p=3$, we have
{\small
\begin{align}
\ket{\phi_3} \equiv\frac{1}{\sqrt{8}} &\Big[\big[3\ket{0}+ 0.293(1{-}i)\ket{1} -i\ket{2} + 1.707(1{+}i)\ket{3}+\ket{4} \nonumber\\ 
&+ 1.707(1{-}i)\ket{5} + i\ket{6} + 0.293(1{+}i)\ket{7} \big]_i\ket{0}_o \nonumber\\
&+ \big[3\ket{0}+ 0.414\ket{1} +1\ket{2} -2.414\ket{3}-\ket{4} -2.414\ket{5}  \nonumber\\
& + \ket{6} + 0.414\ket{7} \big]_i\ket{1}_o + \big[2\ket{0}- (0.707{-}0.293i)\ket{1} \nonumber\\
& -(1{-}i)\ket{2} + (0.707{-}1.707i)\ket{3}+ (0.707{+}1.707i)\ket{5}  \nonumber\\
& - (1{+}i)\ket{6} - (0.707{+}0.293i)\ket{7} \big]_i\ket{2}_o\Big]. 
\label{phi3}
\end{align}
}
Then we can define the reduced density matrix $\rho_p$ as the result when the output register is traced out, that is
\begin{equation}
\rho_p \equiv \Tr_o\big(\ket{\phi_p}\bra{\phi_p}\big).
\label{rhop}
\end{equation}
From using \eq{phi3} and \eq{rhop} we can calculate $\rho_3$ to be
\begin{align}
\resizebox{.95\hsize}{!}{ $10^{-3} \begin{pmatrix}
344 & 11 {+} 5i & 16{+}16i & {-}11{-}27i & 0 & {-}11{-}27i & 16{-}16i & 11{-}5i \\
11{-}5i & 15 & 27{+}11i & {-}31{-}31i & {-}2{-}5i & 22i & 8{-}20i & 9{-}9i \\
16{-}16i & 27{-}11i & 63 & {-}102{-}42i & {-}16{-}16i & 5{+}11i & {-}31i & 8{-}20i \\
{-}11{+}27i & {-}31{+}31i & {-}102{+}42i & 235 & 64{+}27i & 53{+}53i & 5{+}11i & 22i \\
0 & {-}2{+}5i & {-}16{+}16i & 64{-}27i & 31 & 64{+}27i & {-}16{-}16i & {-}2{-}5i \\
{-}11{-}27i & {-}22i & 5{-}11i & 53{-}53i & 64{-}27i & 235 & {-}102{-}42i & {-}31{-}31i \\
16{+}16i & 8{+}20i & 31i & 5{-}11i & {-}16{+}16i & {-}102{+}42i & 63 & 27{+}11i \\
11{+}5i & 9{+}9i & 8{+}20i & {-}22i & {-}2{+}5i & {-}31{+}31i & 27{-}11i & 15
\end{pmatrix}$}
\end{align}
%
%
If one were to measure the input register in the state $\ket{\phi_p}$, the probabilities $\mathcal{P}_p(k)$ of finding the state $\ket{k}$ (for $0 \le k \le 7$) are the values along the diagonal of $\rho_p$. Computing these probabilities for different values of $p$, we have the probability of measuring $\ket{k}$ in the input register for each value or $p$, tabulated in the table \ref{finalprob}.

In effect, each row in table \ref{finalprob} gives the probability distribution of the resulting ket after measuring the input register, for a given period $p$. 
%
%
Practically speaking, this means one can construct the circuit for the function $F_p$, apply the QFT, then measure input register. Repeating the entire process many times, an observed probability distribution, $\overline{\mathcal{P}}(k)$, is obtained and can be compared with the theoretical distribution, $\mathcal{P}_p(k)$, from table \ref{finalprob}. The comparison will help assess the accuracy and effectiveness of the quantum information processing device on which it was implemented. 

\begin{table}[h] \centering
{\small \begin{tabular}{c || c | c | c | c | c | c | c | c }  
 $p$ & $\ket{0}$ & $\ket{1}$ & $\ket{2}$ & $\ket{3}$ & $\ket{4}$ & $\ket{5}$ & $\ket{6}$ & $\ket{7}$  \\ 
  \hline
   1 & 1 & 0 & 0 & 0 & 0 & 0 & 0 & 0  \\
   2 & 0.5 & 0 & 0 & 0 & 0.5 & 0 & 0 & 0  \\
   3 & 0.344 & 0.015 & 0.063 & 0.235 & 0.031 & 0.235 & 0.063 & 0.015  \\
   4 & 0.25 & 0 & 0.25 & 0 & 0.25 & 0 & 0.25 & 0  \\
   5 & 0.219 & 0.059 & 0.125 & 0.19 & 0.031 & 0.191 & 0.125 & 0.059  \\
   6 & 0.188 & 0.125 & 0.063 & 0.125 & 0.188 & 0.125 & 0.063 & 0.125  \\
   7 & 0.156 & 0.147 & 0.125 & 0.103 & 0.093 & 0.103 & 0.125 & 0.147  \\
   8 & 0.125 & 0.125 & 0.125 & 0.125 & 0.125 & 0.125 & 0.125 & 0.125 \\
\end{tabular}
}
\caption{\label{finalprob} The probabilities $\mathcal{P}_p(k)$ of finding $\ket{k}$ after a measurement on the input register of $\ket{\phi_p}$. The column index is $k$ and the row index is $p$.}
\end{table}

To make the model more realistic, one can introduce a depolarizing channel model of noise \cite{king}, via the transform
\begin{equation}
\rho_p \rightarrow \rho_p' = (1-\epsilon)\frac{I}{8} + \epsilon \rho_p,
\label{depol}
\end{equation}
for some parameter $\epsilon$. The value $\epsilon=1$ results in an unchanged (noiseless) state, and $\epsilon=0$ a maximally mixed (totally noisy) state. Under this transformation due to the depolarizing channel, the probability distribution in row $p$ of table \ref{finalprob} will be ``diluted" by the maximally mixed distribution (which corresponds to the $p=8$ row). That is
\begin{equation}
\mathcal{P}_p(\ket{k}) \rightarrow \mathcal{P}_p'(\ket{k})  = \frac{1}{8}(1-\epsilon) + \epsilon\mathcal{P}_p(\ket{k}).
\label{Peps}
\end{equation}
The transformation in \eq{Peps} follows from \eq{depol} together with the realization that the probability distribution $\mathcal{P}_p(\ket{k})$ consists of the diagonal elements of $\rho_p$. The transformed probabilities sum to unity, since $\sum_{k=0}^7\mathcal{P}_p(\ket{k}) = 1$ implies $\sum_{k=0}^7\mathcal{P}_p'(\ket{k}) = 1$.

%

The presence of entanglement between the input and output registers is also of interest. We note that the state $\ket{\psi_p}$ defined in \eq{psip} above is completely separable for $p=1$ and maximally entangled for $p=8$, with intermediate entanglement for $1<p<8$. The state $\ket{\phi_p}$ will have the same entanglement level as $\ket{\psi_p}$, since they are related by a local unitary transformation, the QFT.

Suppose we wish to find some rough measure of entanglement of our pre-measurement state using only the experimentally observed probability distribution, $\overline{\mathcal{P}}(k)$. In this case, we can introduce a new quantity called the \emph{rough separability index}, $S$, defined as the sum of the squares of the observed probabilities:
\begin{equation}
S=\sum_{k=0}^8\overline{\mathcal{P}}(k)^2.
\label{Sdef}
\end{equation}
We calculate $S$ for the theoretical probability distributions $\mathcal{P}_p(k)$ for each period $p$, and tabulate the result in table \ref{roughseparable}.

\begin{table}[h] \centering
\begin{tabular}{c || c | c | c | c | c | c | c | c  }  
   $p$	& 1 &  2  & 3   & 4   & 5   & 6 & 7 & 8\\ \hline
   $S$ 	& 1 & 0.5 & 0.238 & 0.25   & 0.16   & 0.141 & 0.129 & 0.125 \\ 
\end{tabular}
\caption{\label{roughseparable}The rough separability index $S$ for various values of period $p$. The values of $S$ are calculated from the probability distributions in table \ref{finalprob} by summing the squares of the entries for each row.}
\end{table}

We note that $S$ is almost monotonically decreasing when taken as a function of $p$, with just one exception at $p=3$. And since entanglement monotonically increases with $p$, we can take $1-S$ as a rough measure of entanglement created by our circuit. This can act as a proxy for the ability of the constructed circuit to handle entanglement, assuming a simple depolarizing channel noise model.

Using \eq{Peps} and \eq{Sdef}, we find the value of $S$ is affected by the introduction of a depolarization channel model of noise by
\begin{equation}
S \rightarrow S' = \epsilon^2S + \frac{1}{64}(1-\epsilon)(1+15\epsilon).
\label{Sprime}
\end{equation}
Supposing we know the theoretical (noiseless) value of $S$ based on our knowledge of the period $p$, and we measure (noisy) $S'$ after executing the circuit, we can use \eq{Sprime} to find the value of $\epsilon$, and estimate the noisiness of our system.

To sum up, in this section we introduced a simple method to test and validate a nascent quantum technology using some periodic circuit and the QFT. This procedure applied will give us some basic insight into the operation of our test circuit, how well it functions, as well as its handling of noise and entanglement.

\section{Conclusion}
Shor's algorithm is based on probabilistic classical steps as well as a quantum order finding subroutine. The latter has a modular exponentiation step, which in its general form is very difficult to implement with currently realizable quantum technology. Therefore, experimentalists over the last decade or so have resorted to compiled versions of Shor's algorithm, that simplify the modular exponentiation step using known information about the solution.

In this paper, we have extended the method of compiling the modular exponentiation operation, and demonstrated a more efficient method of circuit synthesis than previously used. We have provided the compiled circuits for several semiprimes, and illustrated how the process can be generalized. A simple layer of ``classical compilation" has been discussed as potentially simplifying complicated circuits further. 

An emphasis has been placed on periodicity in the modular exponentiation circuits, and a link to the simple periodic functions in ref. \cite{gamelPeriodic} has been pointed out. Finally, we have presented a simplified procedure to validate the function of such a circuit, including its handling of noise and entanglement.

\section{Acknowledgement}
We would like to thank the Natural Sciences and Engineering Research Council of Canada (NSERC) for funding this research.

\bibliographystyle{unsrt}
\bibliography{ShorBib}

\end{document}